\documentclass[preprint,pre,aps,showpacs]{revtex4}
\usepackage{graphicx}
\begin{document}

\title{Extending the canonical thermodynamic model: inclusion of hypernuclei}

\author{S. Das Gupta}

\affiliation{Physics Department, McGill University, 
Montr{\'e}al, Canada H3A 2T8}

\date{\today}

\begin{abstract}

The canonical thermodynamic model has been used frequently to describe
the disassembly of hot nuclear matter consisting of neutrons and protons.
Such matter is formed in intermediate energy heavy ion collisions.
Here we extend the method to include, in addition to neutrons and protons,
$\Lambda$ particles.  This allows us to include productions of hypernuclei
in intermediate energy heavy ion collisions.  We can easily predict 
average mass number of hypernuclei produced and values of relative 
cross-sections of different $^a_{\Lambda}$z nuclei.  Computation of absolute
cross-section was not attempted at this stage and will require much more
detailed work.

\end{abstract}


\maketitle

\section{Introduction}
The canonical thermodynamic model (CTM) as applied to the nuclear
multifragmentation problem addresses the following scenario.
Assume that because of collisions we have a finite piece of 
nuclear matter which is heated and begins to expand.
During the expansion the nucleus will break up
into many fragments(composites).  In the expanded volume, the 
nuclear interaction between different composites can be neglected
and the long range Coulomb force can be absorbed in a suitable 
approximation.  The partitioning of this hot piece of matter
according to the availability of phase space can be calculated
exactly (but numerically) in CTM.  Many applications of this model
have been made and compared to experimental data \cite{Das1}.  The
model is very similar to the statistical multifragmentation model
(SMM) developed in Copenhagen \cite{Bondorf}.  SMM is more general but requires
complicated Monte-Carlo simulations.  In typical physical situations
the two models give very similar results \cite{Botvina1}.
 
In usual situations, the piece of nuclear matter has neutrons and protons,
that is, it is a two-component system.  Initially CTM was formulated
for one kind of particle \cite{Dasgupta1} and already many interesting
properties like phase transition could be studied.  Subsequent to
the extension of CTM to two kinds of particles
\cite{Bhattacharyya}, many applications of the model to compare with
experimental data were made \cite{Das1,Tsang1}.
The objective of this
paper is to extend CTM to three-component systems.  While this, in general,
is interesting, it can also be useful for calculations in an area
of current interest.  I refer here to the production of hyperons (usually
$\Lambda$) in heavy ion reaction in the 1GeV/A to 2 GeV/A 
beam energy range.  The $\Lambda$'s can get attached to nuclei
turning them into composites with three species.
The conventional thinking is this.  The $\Lambda$
particle is produced in the participating zone, i.e., the region of
violent collisions.  The produced $\Lambda$'s have an extended rapidity
distribution and some of these can be absorbed in the much colder 
specator parts.  These will form hypernuclei.  Those absorbed in the
projectile fragment (PLF) can be more easily studied experimentally because 
they emerge in the forward direction.

This idea has been recently used to study the production of hypernuclei
in a recent paper using the SMM model \cite{Botvina2}.  Our work closely
follows the same physics, however, using a different and what we believe
a much easier prescription.  In addition our focus is different
and we emphasize other aspects.  The question of hypernucleus production
in heavy ion reaction was already looked at in detail more than twenty
years ago \cite{Wakai}.  The authors used a coalescence model.  Both the
break up of a PLF into composites and the absorption of the $\Lambda$
used coalescence.  The coalescence approach has been revived in a much
more ambitious calculation recently \cite{Gaitanos}.  
Intuitively the coalescence model is appealing but
for a satisfactory formulation of the PLF breaking up into many composites
there are many very difficult details
which need to be worked out.  Certainly
there are some points of similarities between the thermodynamic model for
multifragmentation and production of composites by coalescence.  For
the production of the deuteron, the simplest composite, the two models
were compared \cite{Jennings}.  But that study also shows that for a 
heavier fragment (say $^{12}$C) the two routes become impossible to
disentangle from each other.  However, there is at least one argument in
favour of the thermodynamic model (both SMM and CTM).  They have been
widely used for composite production and enjoyed very good success
\cite{Das1,Bondorf}.

The physics ansatz for the calculation reported in the earlier work \cite
{Botvina2} and the prersent work is the same.  The $\Lambda$ particle
(particles) which arrive at the PLF interact strongly with the
nucleons.  Thus fragments can be calculated as in normal prescriptions.
Hypernuclei as well as normal (non-strange) composites will be formed.
The model gives definitive predictions as the following sections will show.
Experiments can vindicate or contradict these predictions.
 
\section{Mathematical Details}
The case ot two-component system (neutrons and protons) have been dealt with
in many places including \cite{Das1}.  The generalisation to
three components is straightforward.
 
The multifragmentation of the system we study has a given number of 
baryons $A$, charges $Z$ and strangeness number $H$.  This will
break up into composites with mass $a$, charge $z$ and $h$ number
of $\Lambda$ particles.  The canonical partition function of the
system $Q_{A,Z,H}$ is given by the following equation.  Once the
partition function is known, observables can be calculated.
\begin{equation}
Q_{A,Z,H}=\sum\prod \frac{(\omega_{a,z,h})^{n_{a,z,h}}}
{n_{a,z,h}!}
\end{equation}
Here $\omega_{a,z,h}$ is the partition function of one composite which
has mass number $a$, charge number $z$ and $h$ hyperons (here $\Lambda$'s)
and $n_{a,z,h}$ is the number of such composites in a given channel.
The sum over channels in eq.(1) is very large and each channel must
satisfy
\begin{eqnarray}
\sum an_{a,z,h} &=& A  \nonumber \\
\sum zn_{a,z,h} &=& Z  \nonumber  \\
\sum hn_{a,z,h} &=& H
\end{eqnarray}
Proceeding further, we have
\begin{equation}
\langle n_{a,z,h} \rangle =\frac{1}{Q_{A,Z,H}}\sum\prod n_{a,z,h}
\frac{(\omega_{a,z,h})^{n_{a,z,h}}}{n_{a,z,h}!}
\end{equation}
which readily leads to
\begin{equation}
\langle n_{a,z,h} \rangle=\frac{1}{Q_{A,Z,H}}\omega_{a,z,h}Q_{A-a,Z-z,H-h}
\end{equation}
Since $\sum a\langle n_{a,z,h}\rangle=A$ we have
\begin{equation}
\sum a\frac{1}{Q_{A,Z,H}}\omega_{a,z,h}Q_{A-a,Z-z,H-h}=A
\end{equation}
which immediately leads to a recuurence relation which can be used to calculate
the many particle partition function:
\begin{equation}
Q_{A,Z,H}=\frac{1}{A}\sum _{a=1}^A a\omega_{a,z,h}Q_{A-a,Z-z,H-h}
\end{equation}
It is obvious other formulae similar to the one above exist:
\begin{equation}
Q_{A,Z,H}=\frac{1}{H}\sum h\omega_{a,z,h}Q_{A-a,Z-z,H-h}
\end{equation}
The above equations are general.  In this paper we do numerical
calculations for the cases $H$=1 and $H$=2.  The composites considered
have either $h$=0 (non-strange composites) or $h$=1.  For the case
$H$=2 this means that in a given channel there will be two
composites each with $h$=1.  A more general treatment would include
composites with two $\Lambda$'s.  To complete the story we need to
write down the specific expressions for $\omega_{a,z,h}$ that we use.

The one particle partition function is a product two parts:
\begin{equation}
\omega_{a,z,h}=z_{kin}(a,z,h)z_{int}(a,z,h)
\end{equation}
The kinetic part is given by
\begin{equation}
z_{kin}(a,z,h)=\frac{V}{h^3}(2\pi MT)^{3/2}
\end{equation}
where $M$ is the mass of the composite: $M=(a-h)m_n+hm_{\Lambda}$.  Here
$m_n$ is the nucleon mass (we use 938 MeV) and $m_{\Lambda}$ is the
$\Lambda$ mass (we use 1116 MeV). For low mass nuclei, we use experimental
values to construct $z_{int}$ and for higher masses a liquid-drop
formula is used.  The neutron, proton and $\Lambda$ particle are taken
as fundamental blocks and so $z_{1,0,0}=z_{1,1,0}=z_{1,0,1}=2$
(spin degeracy).
For deuteron, triton, $^3$He and $^4$He we use $z_{int}(a,z,0)=
(2s_{a,z,0}+1)\exp (-e_{a,z,0}(gr)/T)$ where $e_{a,z,0}$ is the ground
state energy and $(2s_{a,z,0}+1)$ is the experimental spin degeneracy
of the ground stste.  Contributions to the $z_{int}$ from excited states
are left out for these low mass nuclei.  Similarly experimental data
are used for $^3_{\Lambda}$H, $^4_{\Lambda}$H, $^4_{\Lambda}$He and
$^5_{\Lambda}$He.

For heavier nuclei ($h$=0 or 1), a liquid-drop formula is used for 
ground state energy.  This formula is taken from \cite{Botvina2}.
All energies are in MeV.
\begin{equation}
e_{a,z,h}=-16a+\sigma(T)a^{2/3}+0.72z^2/(a^{1/3})+25(a-h-2z)^2/(a-h)
-10.68h+21.27h/(a^{1/3})
\end{equation}
Here $\sigma(T)$ is  temperature dependent surface tension:
$\sigma(T)=18[\frac{T_c^2-t^2}{T_c^2+T^2}]^{5/4}$. A comparative study
of the above binding energy formula can be found in \cite{Botvina2}.
This formula also defines the drip lines.  We include all nuclei
within drip lines in constructing the partition function.

With the liquid-drop formula we also include the contribution to
$z_{int}(a,z,h)$ coming from the excited states.  This gives a
multiplicative factor $=exp (r(T)Ta/\epsilon_0)$ where we have introduced
a correction term $r(T)=\frac{12}{12+T}$ to the expression used in
\cite{Bondorf}.  This slows down the increase of $z_{int}(a,z,h)$  
due to excited states 
as $T$ increases.  Reasons for this correction can be
found in \cite{Bhattacharyya,Koonin} although for the temperature range
used in this paper the correction is not important.

We also incorporate the effects of the long-range Coulomb force
in the Wigner-Seitz approximation \cite{Bondorf}.

We have used eq.(7) to compute the partrition functions.  If the PLF
which absorbs the $\Lambda$ has mass number $A$ and proton number $Z$,
we first calculate all the relevant partion functions for $H$=0 first.
This requires calculating upto $Q_{A,Z,0}$.  We then calculate, for
$H$=1 partition functions upto $Q_{A+1,Z,1}$.  We can then proceed
to calculate for $H$=2 upto $Q_{A+2,Z,2}$ and so on. 

\section{Results for H=1}
We assume one $\Lambda$ is captured in the projectile like fragment (PLF).
The PLF breaks up into various fragments.  In an event one of these fragments
will contain the $\Lambda$ particle, the rest of the fragments will have
$h=0$.  There is also a probability that the $\Lambda$ remains unattached,
i.e., after break-up it emerges as a free $\Lambda$.  There is also
another extreme possibility (this requires very low temperature in the PLF)
that the $\Lambda$ gets attached to the entire PLF which does not break
up.  In such an event the number of composites with $h=0$ is zero. 
The average over all events give the average multiplicity of all
composites, with $h=0$ and $h=1$ (eq.(4)).

We will show results for $\Lambda$ captured by a system of $A=100, Z=40$
and $A=200, Z=80$.  These are the same systems considered in 
\cite{Botvina2}.  The results for $\langle n_{a,z,h} \rangle$ depend
quite sensitively on the temperature and less so on the assumed freeze-out
density.  Except in one case, all the results shown use freeze-out density
to be one-third normal density.  Past experiences have shown 
\cite{Gargi1,Gargi2} that a freeze-out density of one-third normal
density gives better results for disassebly of PLF than, for example,
the value of one-sixth normal density which is more appropriate for the
participating zone.  Again from past experiences, temperatures in the
range 5 to 10 MeV are considered to be appropriate.

In Fig.1 we show results for $A=100, Z=40$ at a low temperature of 
$T$=4 MeV.  In order to display the results easily we sum over 
the charge and plot $\langle n_{a,h} \rangle =\sum_z\langle n_{a,z,h} \rangle$.
Note that for this choice of temperature, the average mass number of
the hypernucleus formed is very high, about 95.  
The multiplicity of non-strange composites is low (1.24).
The average mass of non-strange composite is about 5 and the average charge 
is about 1.7.  We thus have a curious situation.  The non-strange part is a
gas with very few particles and the strange part of matter is a liquid
since in heavy ion physics, a large blob of matter is attributed to be
the liquid part.  While this aspect of 
hybrid liquid-gas co-existence may lead to an
interesting study, our focus here will be the population of hypernuclei.  

For brevity we do not show the population of composites at this temperature
for a system of $A=200, Z=80$.  There are remarkable similarities
in the shapes of the the curves, but the differences are also significant
and the curve for $A=200$ can not be scaled onto the curve for $A$=100.
At higher temperature, however, one can guess the results for $A=200$
knowing, for example, the results for $A$=100.

Fig. 2 shows the graph of $\langle n_{a,h} \rangle$ at 8 MeV temperature
for both $A=100, Z=40$ and the system double its size.  The important feature 
which allows one to scale the results of one system to another is this.
At this temperature the relative population of $\langle n_{a,h} \rangle$
drops off rapidly with $a$ so that the population beyond, say, $a$=40
can be ignored.  For compsites with $h$=1, both the systems $A=100$ and
$A$=200 are virtually the same: for both, $\sum_{a=1}^{40}\langle n_{a,1}
\rangle =1$ and hence it is possible to have the same value of
$\langle n_{a,1}\rangle$ for the the two systems. the graphs for $h=1$
bear this out.
But for $h=0$, $\sum_{a=1}^{a=40}a\langle n_{a,0}\rangle$ have to add up
to different numbers.  For $A=100$ they have to add up to 
$[101-\langle a(h=1)\rangle])$ and for $A=200$ they have to add up to
$[201-\langle a(h=1)\rangle]/$.  The simplest ansatz is
that $\langle n_{a,1}\rangle$ for $A$=200 is larger than the
corresponding quantity for $A=100$ by the ratio
$[201-\langle a(h=1)\rangle]/[101-\langle a(h=1)\rangle]\approx 2$.
Fig. 2 shows this to be approximately correct.

For our model to be physically relevant, we expect the yield
$\langle n_{a,h} \rangle$ to be proportional to $\sigma (a,h)$ although
the model, at the moment, is not capable of providing the value of
the proportionality constant.  The average value:$\langle a(h=1)\rangle$=
$\sum_a a\langle n_{a,1} \rangle/\sum_a\langle n_{a,1}\rangle=
\sum_a a\langle n_{a,1} \rangle$ is a useful quantity and is
predicted to be the average value  of the mass number of the 
hypernuclei measured in experiment.  This is plotted in Fig.3 as 
a function of temperature both for $A$=100 and $A=200$  calculated
at one-third the nuclear density (graphs labelled 1 and 3 respectively).
Curves labelled 2 and 4 refer to the cases when $H$=2 and we deal with 
them in the next section.
In the figure we also plot the value of 
$\langle a(h=1) \rangle$ if this is calculated at a lower one-sixth
normal density for $A$=100 (graph labelled 5).
Several comments can be made.  Assuming that the
PLF temperature is in the expected 6 MeV to 10 MeV range the average 
mass number of hypernuclei should be in 20 to 7 range.  Secondly this 
value is insensitive to the PLF mass number so long as it is reasonably
large.  We can also use the graph to state
that if the temperature is above 6 MeV the grand canonical model
can give a dependable estimate but if the temperature is significantly lower,
say 5 MeV, grand canonical calculation can be in significant error.
As expected if a lower value for the freeze-out density is used 
the predicted value for $\langle a(h=1) \rangle$ is lowered.

A more detailed plot of yields for $^a_{\Lambda}z$ for 
$z$ in the range 1 to 6 and all relevant $a$'s is given in Fig.4. 
There are two curves for each $z$.  Let us concentrate on the lower curves.
These belong to the case considered here, i.e., $H$=1.
Although we have drawn this this for $A=100, Z=40$ for $T$=8 MeV
it is virtually unchanged  for $A=200, Z=80$.  The reasons were 
already given.  These plots provide a very stringent test of the model
as these yields are proportional to experimental cross-sections.

\section{Results for H=2}
We consider now $A=100, Z=40$ and $A=200, Z=80$ but these systems have captured
two $\Lambda$'s rather than one.  How do we expect the results to change?
Fig.5 compares the yields of the composites with two $\Lambda$'s entrapped
in $A=$100 at 8 MeV with the already studied case of one $\Lambda$
in $A$=100 at 8 MeV.  For $H$=2 the number of hypernuclei (and also
the number of free $\Lambda$'s) is doubled with only very small changes
in the number of non-strange composites.  We can understand why this
happens following a similar chain of arguments as presented in the
previous section.  The reason for this correspondence is that at
temperature 8 MeV there are only insignificant number of composites
beyond $a$=40.

The average value of $\langle a(h=1)\rangle$ as a function of 
temperature is shown in Fig.3 for $H$=2.  Curve 2 is for $A$=100, $H$=2
and curve 1 is for $A$=100, $H$=1.  As explained above, for $T>7$
tha average value $<a(h=1)>$ will be very close but at lower temperature
(i.e.,$T=4 MeV$) the situation is very different.  For $H$=1 there
is a very large hypernucleus containing most of the nucleons (Fig.1)
but for $H$=2 there are two hypernuclei thus they will together share
the bulk of the nucleons.  Thus the average value of $\langle a(h=1)
\rangle$ will drop to about half the value obtained for $H$=1.
In Fig.3 curve 4 is for $H$=2 in a system with $A$=200, curve 3 is for
$H$=1 in a system with $A$=200.

\section{Discussion}
We have given a detailed description of what happens once the PLF
captures the produced $\Lambda$ particle.  How $\Lambda$'s are 
produced in the violent collision zone and the probability of arrival
both temporally and positionally at PLF are not described here.
This will depend strongly on the experiment: for example, the
case of say,$^{197}$Au hitting $^{12}$C will have to be treated
differently from that of Sn on Sn collisions.  We hope to embark  
upon this aspect in future.  We have looked at statistical
aspects only.   This can be investigated more easily using
the canonical thermodynamic model.

The calculations here looked at productions of hypernuclei in the
PLF.  The technique can also be applied in the participant zone.
In the participant zone the temperature will be higher.
Also the freeze-out density is expected to be lower.  As an example
if we use freeze-out density 1/6-th of the normal density, 
$A=100, Z=40$ and temperature
18 MeV the average value of $a$ for $h$=1 is 2.55.  We produce more
single $\Lambda$'s than hypernuclei.  Heavier hyparnuclei are not
favoured at high temperature.
   
\section{Acknowledgement}
The author is indebted to A. Botvina for drawing his attention to
the topic of this paper.  He also thanks him for many discussions.
This work is supported by Natural Sciences and Engineering Research
Council of Canada.

\begin{figure}
\includegraphics[width=5.5in,height=4.5in,clip,angle=-90]{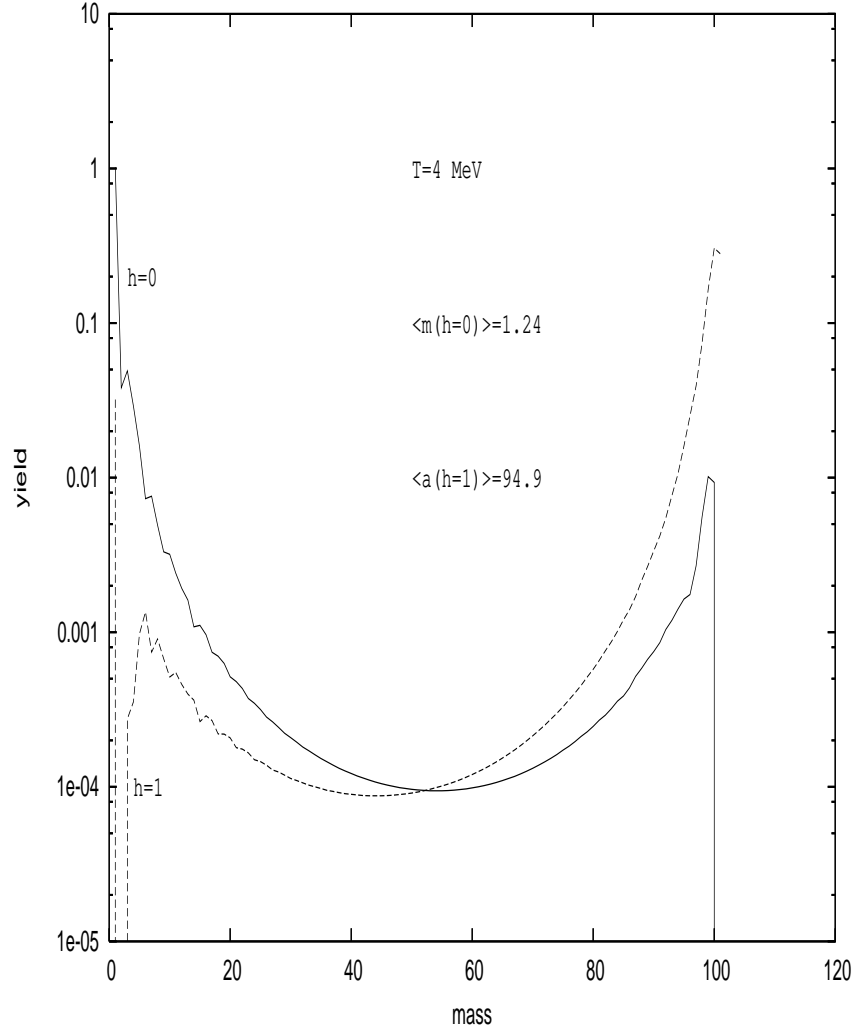}
\caption{For $A=100, Z=40$ CTM results for average yields of composites
$\langle n_{a,h}\rangle$ for $h$=0 (solid line) and $h$=1 (dashed line).
The freeze-out density is 1/3 of normal density.  The case shown uses
a temperature of 4 MeV.  Note that the yields first drop off but then
rise again (specially for $h$=1 case).  This is a case of liquid-gas
co-existence.
The average value of multiplicity for non-strange composites
$<m(h=0)>$ is 1.24
and the average mass number $<a(h=1)>$ of composites with one $\Lambda$
is 94.9.  These values are strongly dependent on the temperature as shown
in Fig2.}
\end{figure}
\begin{figure}
\includegraphics[width=5.5in,height=4.5in,clip,angle=-90]{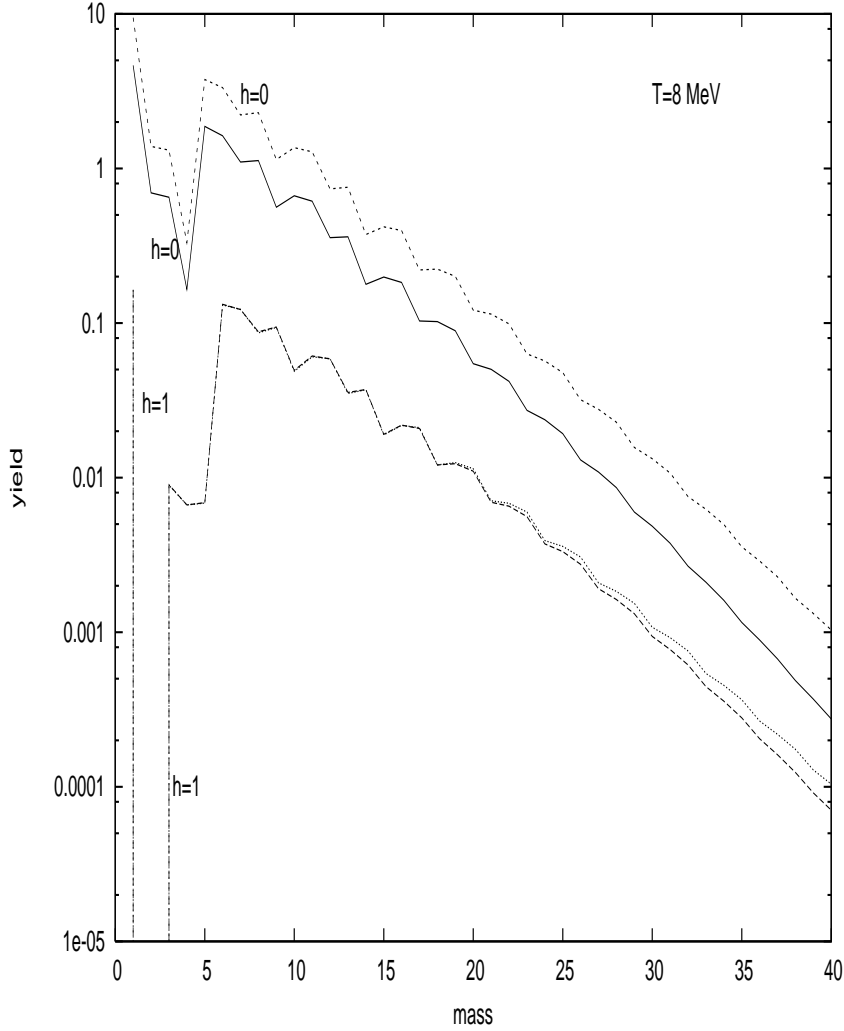}
\caption{ Again we plot average yields $\langle n_{a,h} \rangle$
at 1/3-rd normal density but now at temperature $T$=8 MeV.  We show
results for both $A=100, Z=40$ and $A=200, Z=80$.  Note that 
the pattern of yields have completely changed from that at
temperature 4 MeV.  The yields fall off rapidly with $a$.  
They do not further rise again.  Under such
condtions the population of $h$=1 composites will be remarkably
same for both $A$=100 and 200.  This is explained more fully in the
text.  However, the yields of $h$=0 for $A$=100 (shown by a solid line)
and $A$=200 (dashed curve) can be expected to differ by roughly a 
constant factor.}
\end{figure}
\begin{figure}
\includegraphics[width=5.5in,height=4.5in,clip,angle=-90]{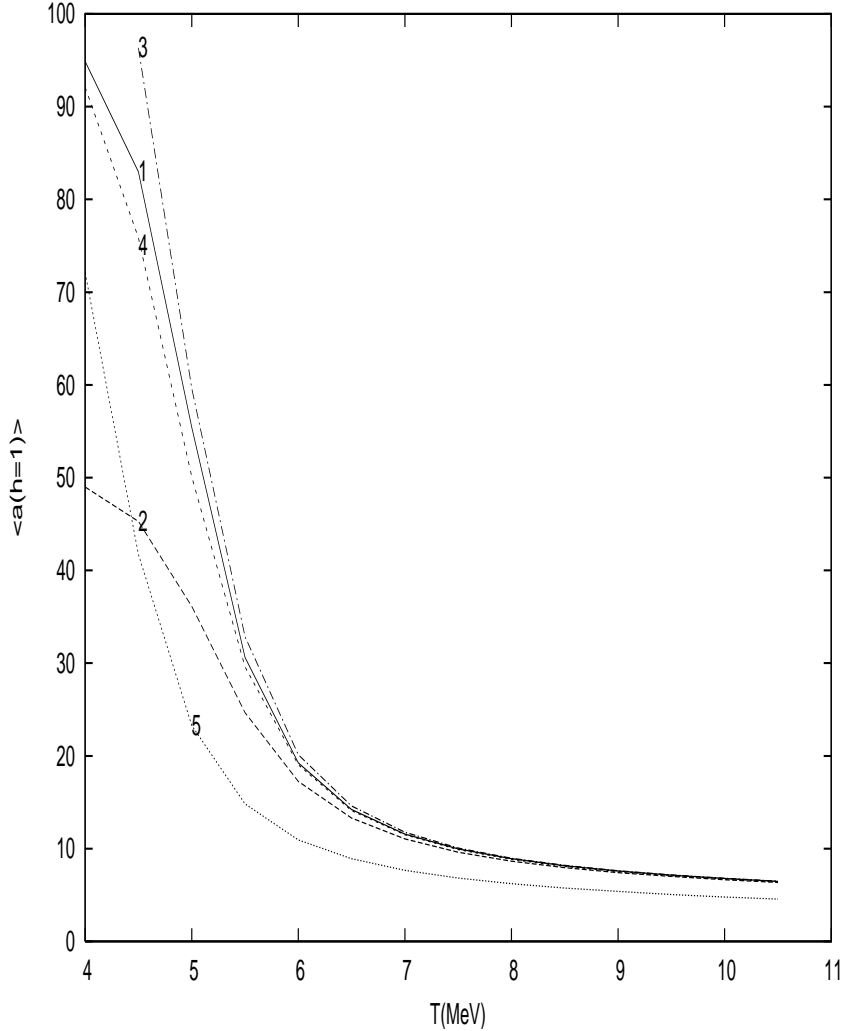}
\caption{The avarage mass number of hypernuclei for
five different scenarios.  If the average mass
number of hypernuclei is measured in heavy ion reactions 
these numbers can be directly compared.  The average
mass is a function of the temperature in the PLF.  
The temperature range of 6 to 10 MeV might be most relevant.
Curve 1 is for PLF with $A=100, Z=40,H=1$ (one $\Lambda$ absorbed).
The freeze-out density is one-third
normal density.  Curve 5 is the same system but a smaller freeze-out
density (1/6 th normal density): shows how the average
value $\langle n_{a,1}\rangle$ changes with freeze-out density.
Except for curve 5 all other curves
use 1/3-rd normal density.  Curve 3 has $A=200, Z=80,H$=1.
Note that except at low temperature
the average value $\langle n_{a,1}\rangle$ does not distinguish
much between systems with $A$=100 and $A$=200.  Curves 2 and 4 are
drawn to show how results differ when two $\Lambda$'s rather than 
one are absorbed by the PLF.
Curve 2 has $A=100, Z=40, H=2$. Curve 4 has $A=200, Z=80, H$=2.
Note that above 6 MeV temperature
curves computed for freeze-out density 1/3-rd give very 
similar result.  For low temperatures the $H$=2 gives
about half the value from what is obtained for $H$=1.  See text for
an explanation.}
\end{figure}

\begin{figure}
\includegraphics[width=5.5in,height=4.5in,clip,angle=-90]{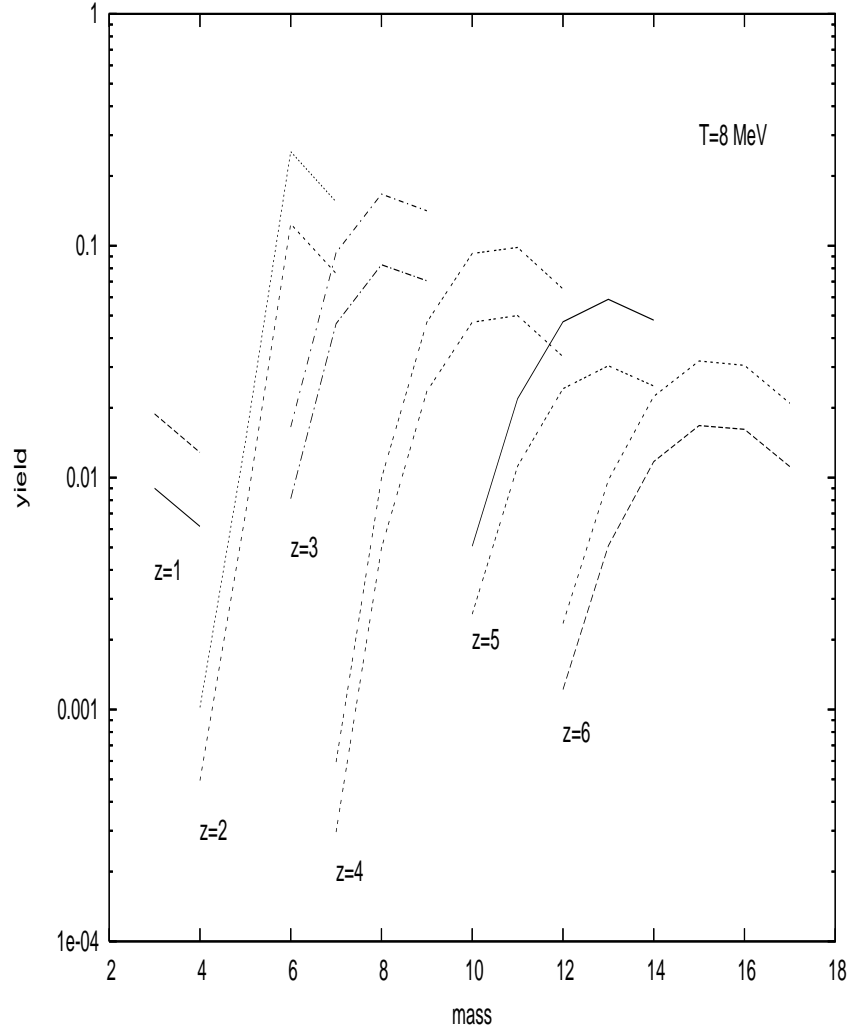}
\caption{For $A=100, Z=40$ we plot $\langle n_{a,z,1}\rangle$
for a range of $z$'s and
relevant $a$'s for $H$=1 (lower curves) and $H$=2 (upper curves). 
Calculations done at 1/3-rd normal nuclear density and temperature 8 MeV.
These yields should be proportional to the measured cross-sections
$\sigma(a,z,1)$.  For $A=200, Z=80$, the results are almost the same.}
\end{figure}
\begin{figure}
\includegraphics[width=5.5in,height=4.5in,clip,angle=-90]{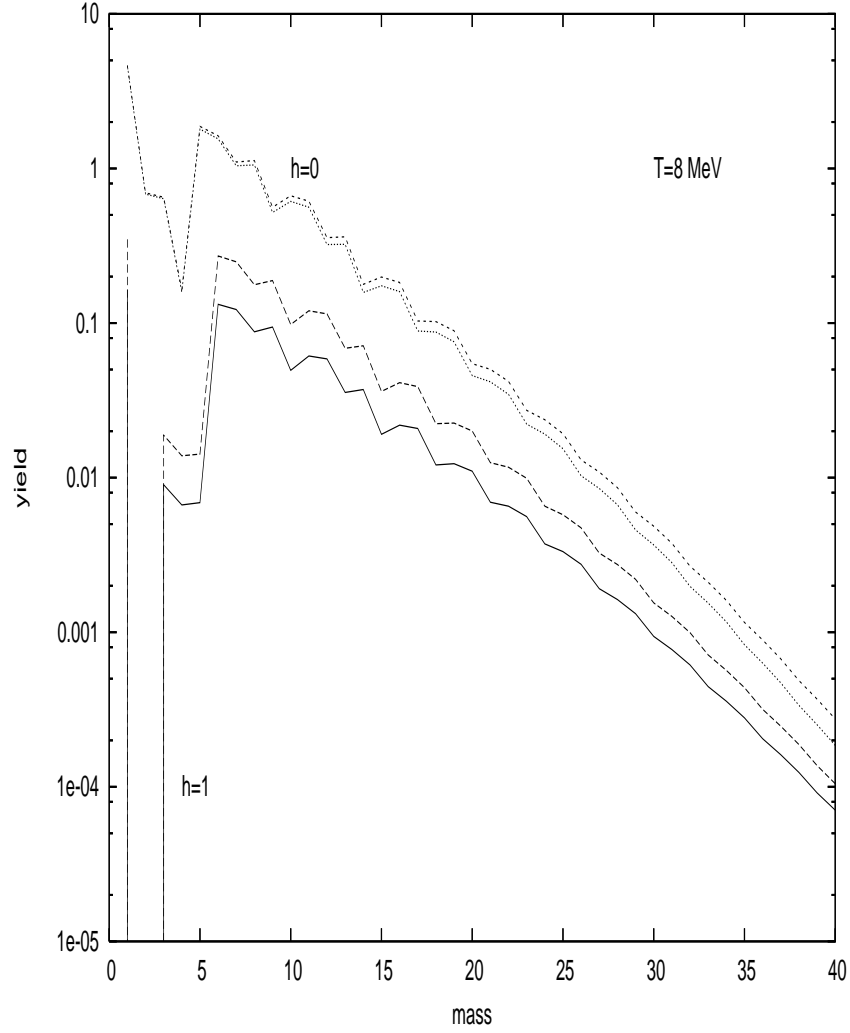}
\caption{Plots of $\langle n_{a,h}\rangle$ for $h$=1 and $h$=0.
The absorbing system is $A=100, Z=40$.  We show results for both
$H$=1 and 2. For $H$=2 the yields of hypernuclei are nearly a factor
of 2 higher than the case with $H$=1.  Populations of normal
composites (h=0) do not alter much between the two cases.}

\end{figure}

\end{document}